\documentclass[aps,nofootinbib,pra,notitlepage,twocolumn,floatfix,longbibliography]{revtex4-1}

\usepackage{amsfonts,amsmath,amssymb,amsthm}
\usepackage{acronym,array,bm,color,dcolumn}
\usepackage{graphicx,nomencl,revsymb4-1,upgreek,url}
\usepackage{hyperref}
\usepackage{subfigure}
\usepackage{tikz}
\usepackage{tkz-graph,tkz-berge}
\hypersetup{colorlinks=true, pdfauthor=Kevin C. Young, pdftitle=}

\newcommand{\ket}[1]{\ensuremath{\left\vert{#1}\right\rangle}}

\newcommand{\abs}[1]{\left\vert #1 \right\vert}
\newcommand{\expect}[1]{\ensuremath{\left\langle{#1}\right\rangle}}

\newcommand{\ident}{\openone}

\newcommand{\cart}{\,\raisebox{0.5pt}{$\mathstrut \scriptstyle \square$}\,}

\newcommand{\rms}{{\mathrm{RMS}}}
\newcommand{\adj}{{\mathrm{Adj}}}

\renewcommand{\to}{\,..\,}
\definecolor{myred}{RGB}{146,19,11}
\definecolor{mygrey}{RGB}{174,157,140}
\definecolor{myblue}{RGB}{21,63,102}
\newcommand{\gcirc}[2][mygrey,fill=mygrey]{\tikz[baseline=-0.5ex]\draw[#1,radius=#2] (0,0) circle ;}%
\newcommand{\bcirc}[2][myblue,fill=myblue]{\tikz[baseline=-0.5ex]\draw[#1,radius=#2] (0,0) circle ;}%
\newcommand{\bog}{\substack{\bcirc{1.5pt}\\ \gcirc{1.5pt}}\,}
\newcommand{\gob}{\substack{\gcirc{1.5pt}\\ \bcirc{1.5pt}}\,}
\newcommand{\gog}{\substack{\gcirc{1.5pt}\\ \gcirc{1.5pt}}\,}
\newcommand{\bob}{\substack{\bcirc{1.5pt}\\ \bcirc{1.5pt}}\,}

\newcommand{\bes} {\begin{subequations}}
\newcommand{\ees} {\end{subequations}}

\begin{document}

\title{Adiabatic quantum optimization with the wrong Hamiltonian}

\author{Kevin C.~Young}
	\email[Electronic address: ]{kyoung@sandia.gov}
	\affiliation{Scalable \& Secure Systems Research (08961),\\
	Sandia National Laboratories, Livermore, CA 94550, USA}
	\author{Robin Blume-Kohout}
	\affiliation{Advanced Device Technologies (01425),\\
	Sandia National Laboratories, Albuquerque, NM 87185, USA}
\author{Daniel A. Lidar}
	\affiliation{Departments of Chemistry, Electrical Engineering, and Physics,
	and the Center for Quantum Information Science \& Technology,
	University of Southern California, Los Angeles, CA 90089, USA}

\begin{abstract}
Analog models of quantum information processing, such as adiabatic quantum computation and analog quantum simulation, require the ability to subject a system to precisely specified Hamiltonians.  Unfortunately, the hardware used to implement these Hamiltonians will be imperfect and limited in its precision.  Even small perturbations and imprecisions can have profound effects on the nature of the ground state.  Here we consider an imperfect implementation of adiabatic quantum optimization and show that, for a widely applicable random control noise model, quantum stabilizer encodings are able to reduce the effective noise magnitude and thus improve the likelihood of a successful computation or simulation.  This reduction builds upon two design principles: summation of equivalent logical operators to increase the energy scale of the encoded optimization problem, and the inclusion of a penalty term comprising the sum of the code stabilizer elements. We illustrate our findings with an Ising ladder and show that classical repetition coding drastically increases the probability that the ground state of a perturbed model is decodable to that of the unperturbed model, while using only realistic two-body interaction.  Finally, we note that the repetition encoding is a special case of quantum stabilizer encodings, and show that this in principle allows us to generalize our results to many types of analog quantum information processing, albeit at the expense of many-body interactions.
\end{abstract}

\maketitle

\section{Introduction} 
\label{sec:introduction}\noindent
Analog paradigms of quantum information processing (QIP) have been extensively studied as possible alternatives to the more standard, but resource-intensive, circuit model, in which a desired quantum evolution is decomposed into a sequence of discrete unitary operators drawn from a small, but universal set \cite{Nielsen:book}.  This digitization enables the celebrated {fault tolerance theorem} \cite{Aliferis:05,Aharonov:08}, but the overhead required for error correction can be demanding.  In an attempt to avoid these overheads, analog QIP, as exemplified by adiabatic quantum computation (AQC) \cite{FarhiAQC:00} or analog quantum simulation \cite{Buluta:2009fk}, proceeds instead by imposing a (possibly slowly time varying) Hamiltonian on the system and measuring system observables after some time.  Working at the Hamiltonian level offers a number of potential advantages, including reduced demands on the control systems and possible robustness to certain noise processes \cite{PhysRevA.65.012322,PhysRevLett.95.250503}.  Unfortunately, provable statements about the performance of analog QIP are few and far between \cite{jordan2006error,PhysRevLett.100.160506}, and there is not yet a clear path to an analog version of the fault tolerance theorem, nor a compelling reason to believe that one might exist \cite{Young:13}.

A number of previous works \cite{jordan2006error,PhysRevLett.100.160506,PhysRevA.86.042333,Young:13,PAL:13,Ganti:13} have investigated the susceptibility of analog QIP to environmental noise and have shown quantum stabilizer codes to be particularly useful for suppressing, and possibly correcting, the resulting errors.  But environmental couplings are not the only source of error in analog QIP.
We may ask, for example, what happens when the desired Hamiltonian is specified to a precision higher than the physical apparatus is capable of implementing?  If the system is tuned near a critical point, for instance, even small imprecision may be sufficient to induce a phase transition.  The ground state of the physical system would then be drastically different from that of the desired Hamiltonian and would, in the case of an AQC, provide an incorrect answer. In this work, we show that stabilizer encodings, in addition to their usefulness in suppressing environmental couplings, are further capable of combating these Hamiltonian implementation errors.

Building on an idea introduced in \cite{PAL:13} for suppression of environmental noise, we present an encoding technique which, for an experimentally plausible model of random control errors, is capable of drastically improving performance.  We begin by considering an AQC designed to find the ground states of programmable Ising models, i.e., perform a quadratic unconstrained binary optimization (QUBO), in the presence of imprecise couplings.  We show that the effective magnitude of imprecision errors may be reduced by a simple repetition encoding.  Finally, we show that this encoding technique functions by leveraging a seldom used property of quantum stabilizer codes: that they provide many different ways to encode any given term in the Hamiltonian.  It is this redundancy which yields a robustness to imprecision errors and increases the likelihood of a successful computation/simulation.  This observation allows us to generalize the encoding technique to problems beyond QUBO. 

The structure of this paper is as follows. In Section~\ref{sec:problem_statement} we define the error model and give an example of a QUBO instance with randomly perturbed couplings. In Section~\ref{sec:protecting_qubo_with_repetition_encoding} we first discuss the general encoding scheme and then apply it to the previous example of a faulty QUBO instance. In Section~\ref{sec:formulation_in_terms_of_stabilizer_codes} we generalize the result to arbitrary stabilizer codes. We conclude in Section~\ref{sec:discussion}.

\section{Problem statement} 
\label{sec:problem_statement}
\noindent
\subsection{Error model}
Solving the QUBO problem with the quantum adiabatic algorithm begins with a system of qubits in the easily prepared ground state of an applied, separable Hamiltonian, such as the transverse field Hamiltonian $H(0) = - \sum_i X_i$, where $X_i$ is the Pauli X operator on the $i^{\rm th}$ qubit, and the sum is taken over all of the qubits in the system.  The Hamiltonian is then slowly, or \emph{adiabatically}, varied so that at the final time $t_f$ it is equal to a predetermined \emph{problem Hamiltonian} $H(t_f)$.  The ground state of this problem Hamiltonian encodes the solution to our problem. In the case of QUBO, the final Hamiltonian is chosen by exploiting a simple formal equivalence between the logical QUBO problem and a physically realizable, generalized, longitudinal-field Ising model on a graph.  As long as the interpolation proceeds sufficiently slowly and there are no additional couplings or noise, the adiabatic theorem guarantees that the system will remain in its ground state throughout, up to an error that can be made arbitrarily small \cite{Teufel:book,Jansen:07,lidar:102106,Wiebe:12}. Measurements of the final state will then reveal the solution to the QUBO problem.  

\begin{figure}
	\begin{center}
	\centerline{\hbox{\hspace{1.375pt}\includegraphics[scale=0.55]{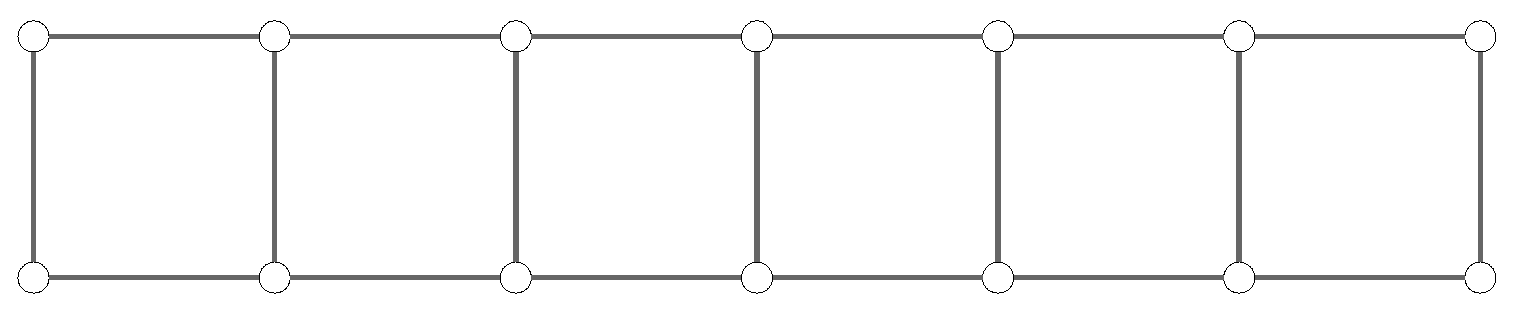}}}\vspace{.3cm}
	\centerline{\hbox{\includegraphics[scale=0.55]{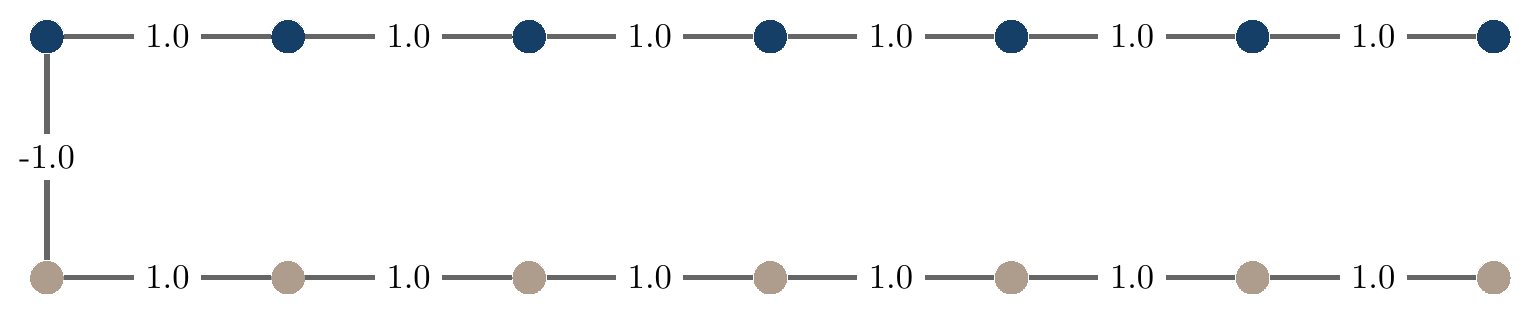}}}\vspace{.3cm}
	\centerline{\hbox{\hspace{.55pt}\includegraphics[scale=0.55]{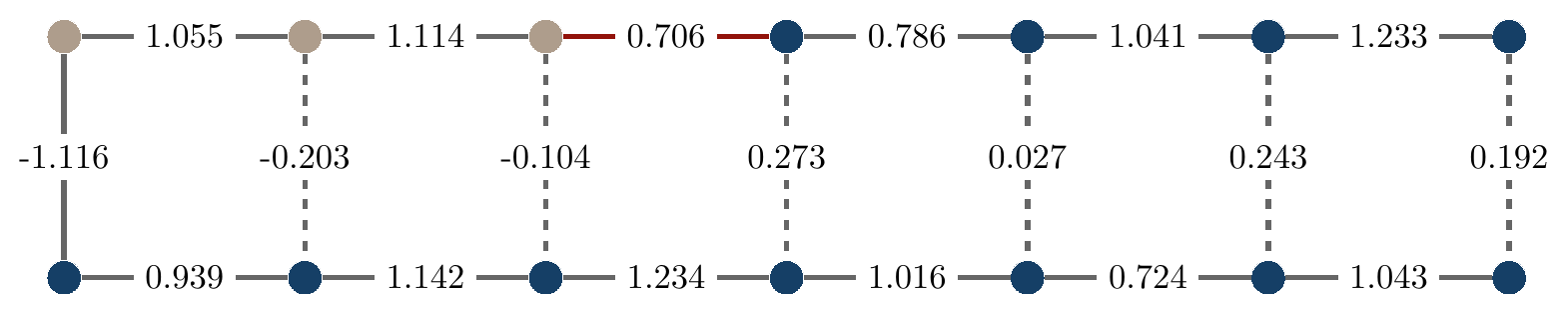}}}
	\caption{Sample QUBO instance susceptible to errors. \\
	(Top) \textbf{Hardware graph} - 
		The qubit locations are shown in white circles and the allowable couplings with solid lines. \\
	(Middle) \textbf{Desired model} - 
		Two parallel Ising chains linked with a single antiferromagnetic coupling.  
		Nearest neighbor couplings of the form $-\alpha Z_i Z_j$, with $\alpha$ 
		labeling the edges ($\alpha=0$ couplings not shown).  
		Also indicated is the spin configuration for one of the two degenerate 
		ground states of the model with blue (tan) circles representing qubit state \ket{0} (\ket{1}). \\
	(Bottom) \textbf{Erred model} - The  model in the presence of noise on the 
		couplings possesses drastically different ground states.  Dashed lines 
		indicate unintended couplings, while red lines indicate violated links.  
		In this case, sufficiently many of the unintended couplings are satisfied to 
		overcome the energetic cost of violating the indicated link (red, labeled ``0.706'').}
	\label{fig:error}
	\end{center}
\end{figure}

While specification 
errors in the intermediate Hamiltonian can cause serious problems (e.g., if they result in a closure of the minimum energy gap and lead to non-adiabatic transitions), henceforth we only consider the problem Hamiltonian $H(t_f)$. Of course, even if the adiabatic interpolation proceeds without error, if the final Hamiltonian is implemented incorrectly then we will have solved the wrong problem. Because we are focusing on QUBO problems, the final Hamiltonian $H(t_f)$ takes the form of a generalized Ising model on a graph:
\begin{equation}
\label{eq:ham}
	H_{\rm Q} = \sum_{\left\langle i,j \right\rangle \in \mathcal{E}} J_{ij} Z_i Z_j + \sum_{i\in\mathcal{V}} h_i Z_i,
\end{equation}
where the first sum is over all links in the edge set, $\mathcal{E}$, of the graph and the second sum is over all vertices in the vertex set, $\mathcal{V}$.  We shall consider the graph structure to be hardware constrained.  That is, the edge set dictates all links along which couplings \emph{may} exist, though any may be chosen to be zero for any particular problem instance, see Fig.~\ref{fig:error} (top).  The Pauli operators, $Z_i$, have eigenvalues $\pm 1$, and the energy scales are presumed bounded, $\abs{J_{ij}}, \abs{h_i} \le E_{\rm max} $.

We model imprecision in the couplings by assuming that the system is additionally subject to an error Hamiltonian,
\begin{equation}
\label{eq:dham}
	\Delta H = \sum_{\left\langle i,j \right\rangle \in \mathcal{E}} \epsilon^{(J)}_{ij} Z_i Z_j + \sum_{i\in\mathcal{V}} \epsilon^{(h)}_i Z_i.
\end{equation}
As before, nonzero error couplings can exist only on links in the hardware graph.  We make a number of reasonable assumptions about this Hamiltonian, restricting our discussion to situations in which systematic errors have been eliminated by experimental calibration, and the coupling errors are 
\begin{enumerate}
	\item \emph{randomly distributed with zero mean}, i.e., $\expect{\epsilon} = 0$
	\item \emph{uncorrelated}, i.e., independent of one another and of the values of $J_{ij}$ and $h_i$
	\item \emph{small}, i.e., significantly smaller than the maximum intended couplings: $\epsilon_\rms \equiv \expect{\epsilon^2}^{1/2} \ll E_{\rm max}$.
\end{enumerate}

A feature of this error Hamiltonian is that it commutes with the final-time QUBO problem Hamiltonian, though not necessarily with any of the intermediate-time Hamiltonians. These errors therefore only affect the \emph{energies} of the final eigenstates, which are simply computational basis states as the final Hamiltonian is a sum of $Z$-like Pauli operators.  Nonetheless, the errors may sufficiently perturb these energies that the ground state of $H^\prime=H_{\rm Q}+\Delta H$ is no longer equal to the ground state of the unperturbed problem Hamiltonian, $H_{\rm Q}$.  

\subsection{Example of a faulty QUBO instance}
Figure~\ref{fig:error} illustrates a simple QUBO problem instance which is particularly susceptible to imprecisions in the coupling terms.   The Hamiltonian for this problem is $H_Q = -\sum_{ij}J_{ij}Z_i Z_j$, with $J_{ij}=\pm 1$ as indicated in Fig.~\ref{fig:error} (middle).  This instance consists of two ferromagnetic Ising chains linked antiferromagnetically at one end, with $J_{ij}=0$ for all other cross links on the ladder.  This Hamiltonian remains unchanged upon flipping all qubits, so the the ground space of this problem is doubly degenerate.  We label the states graphically, with blue and tan dots representing qubit states $\ket 0$ and $\ket 1$, respectively.  In this notation, the ground space of the problem Hamiltonian is spanned by the states  $\ket{\bog\bog\bog\bog\bog\bog\bog}$ and $\ket{\gob\gob\gob\gob\gob\gob\gob}$, with the top and bottom rows representing the corresponding rows in Fig.~\ref{fig:error}.
Coupling errors on all links in the hardware graph are drawn uniformly from the range $[-0.3,0.3]$ and we are neglecting on-site errors (so $\epsilon^{(h)}_i = 0$).  The new ground states were determined by exploiting a formal equivalence with Weighted MAX-2-SAT (discussed in Appendix~\ref{sec:encoding_as_max_sat}) and solved using the freely available SAT solver, AKMAXSAT \cite{Alsinet:2005wy}.  Figure~\ref{fig:error} (bottom) shows an example with these erred couplings which, though weak, are sufficient to change the ground states to $\ket{\bog\bog\bog\gog\gog\gog\gog}$ and $\ket{\gob\gob\gob\bob\bob\bob\bob}$.

As the ladder grows longer, the problem becomes increasingly susceptible to weak errors.  This may be understood by considering the energy cost associated with violating a single problem link, as occurs in the example shown in Fig.~\ref{fig:error} (bottom).  In this case, this energy cost was offset by energy savings due to satisfying several of the erroneous (dashed) crosslink couplings.  If the total number of qubits in the system is $2N$, then the total energy associated with the crosslinks is approximately $\sqrt{N}\epsilon_\rms$.  If this is larger than the energy associated with a single problem link, $\sqrt{N}\epsilon_\rms \ge E_{\rm max}$, then it is likely that the erroneous couplings will cause a change in the ground state.  This $\sqrt{N}$ increase in failure probability appears in our Monte Carlo (MC) experiments, as we discuss in detail below.

In what follows we will assume that the adiabatic algorithm proceeds as desired, so that the final state is the ground state of the perturbed Hamiltonian, $H^\prime$.  We shall therefore be interested in encoding constructions which suppress the effects of $\Delta H$.

\begin{figure}[t]
	\begin{center}
	\includegraphics[width=\columnwidth]{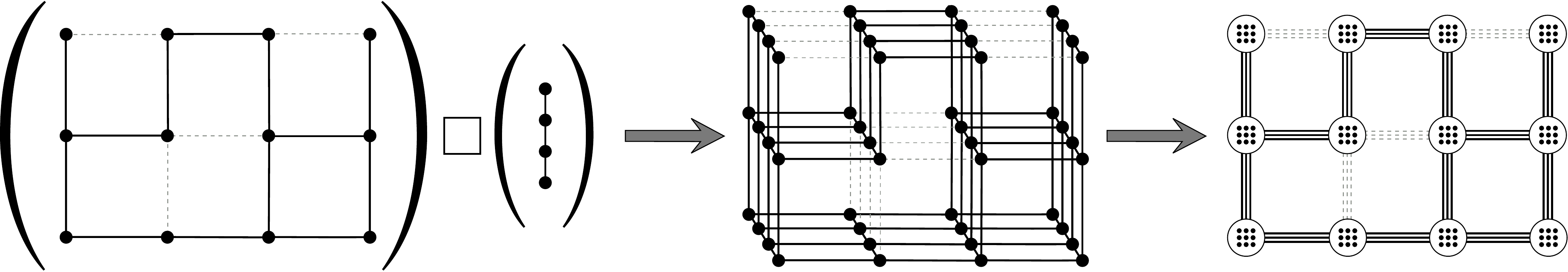}
	\caption{Encoding a generic QUBO problem with a repetition code.  The QUBO connectivity graph (extreme left) is replicated $K$ times and linked at equivalent vertices with ferromagnetic interactions (shown here with linear connectivity for ease of presentation, but higher connectivity is preferable).  This is equivalent to taking a Cartesian product of the original QUBO graph with an Ising ferromagnet (middle).  In the code space of the repetition code, all the spins in each code block will align and may be treated as a single dynamical variable under a renormalized Hamiltonian (right).}
	\label{fig:rep}
	\end{center}
\end{figure}

\section{Protecting QUBO with repetition encoding} 
\label{sec:protecting_qubo_with_repetition_encoding}
\noindent
\subsection{General protection scheme}
To protect the system, we encode each logical qubit in the problem Hamiltonian as $K$ ferromagnetically coupled physical qubits with coupling constant $J_F>0$. This idea was recently used to demonstrate a fidelity improvement in experimental quantum annealing \cite{PAL:13}, but the effect of the error Hamiltonian $\Delta H$ was not considered.  The encoding introduces a new label, $k$, indexing the physical qubit within each logical qubit, so that $Z_i \mapsto Z_{i,k}$.  The connectivity graph of the encoded system is endowed with a Cartesian product structure, so that if $G = (\mathcal{E},\mathcal{V})$ is the graph of the original system and $F= (\mathcal{E}_F,\mathcal{V}_F)$ is the Ising ferromagnet used to encode the logical qubits, then the new graph is $G\cart F$.  In terms of the adjacency matrices, 
\label{eq:cart}
$\adj(G \cart F) = \adj(G)\otimes I_F + I_G\otimes \adj(F)$ (see \cite{alg-graph:book}, p.~52).
Here $I_G$ ($I_F$) is the identity matrix of dimension equal to the dimension of $G$ ($F$).  See Fig.~\ref{fig:rep}. 

Including error terms, the Hamiltonian after encoding becomes,
\begin{align}
	\overline{H^\prime}
		\notag
		=& \sum_{\left\langle i,j \right\rangle \in \mathcal{E}}\sum_{k\in \mathcal{V}_F} 
			\left(J_{ij}+\epsilon_{ij,k}^{(J)}\right) Z_{i,k} Z_{j,k} \\
		\notag
		&+ \sum_{i\in\mathcal{V}}\sum_{k \in \mathcal{V}_F}\left(h_i+\epsilon^{(h)}_{i,k}\right) Z_{i,k} \\
		\label{eq:hnew}
		&+ \sum_{i\in\mathcal{V}}\sum_{\left\langle k,l \right\rangle \in \mathcal{E}_F}  \,(-J_F +\epsilon_{i,kl}^{(J)}) Z_{i,k}Z_{i,l}
\end{align}
If the connectivity of the ferromagnetic Ising blocks is higher than the logical qubit connectivity, and the perturbations are weak on the scale of the Ising couplings, then 
with high probability, when the overall system is in its ground state, the state of each logical qubit will lie in its code subspace.  
This can be understood by a simple energy accounting argument.  This encoding repeats the model $K$ times and links the model copies together.  Now suppose that the error couplings conspire so that the ground state of one of these repeated models is different from the others.  The energy benefit associated with flipping $M$ spins to satisfy the error Hamiltonian $\Delta H$ will be, on average, $M\epsilon_\rms$.  However, the energy \emph{cost} associated with these $M$ spins violating the ferromagnetic couplings is approximately $d M J_F$, where $d$ is the degree of the encoding graph, and $J_F\simeq E_{\rm max}\gg\epsilon_\rms$.  The energy cost is higher than the energy benefit, so the ground state of the perturbed model is likely to lie in the code space.  Here we have ignored errors in the encoding couplings, because weak perturbations cannot cause a change in the ferromagnetic character of the Ising model ground state.

To demonstrate how the encoding suppresses the error Hamiltonian, it will be useful to define the code space projector, where each copy of the Ising ferromagnet is in a ground state, as:
\begin{align}
	P &= \frac{1}{2^K}\prod_{i\in\mathcal{V}}\left(\prod_{k\in\mathcal{V}_F}(1-Z_{i,k}) 
		+ \prod_{k\in\mathcal{V}_F}(1+Z_{i,k})\right) 
		\label{eq:projector}
\end{align}
In the code space, many physical qubit operators will act identically, so may be replaced by a single logical operator describing all the spins in the logical qubit,
\begin{equation}
	P Z_{i,k} P = P Z_{i,1} P \equiv P Z_{i} P .
	\label{eq:PZP}
\end{equation}
Performing this substitution and summing over the physical qubit index, $k$, gives, up to an overall energy shift,
\begin{align}
\label{eq:hnewB}
P\overline{H^\prime}P 
			=  P\left( K H_Q  + \sqrt{K}\sum_{\left\langle i,j \right\rangle \in \mathcal{E}}  
				 \epsilon^{(J)}_\rms Z_i Z_j \right)P .
	\end{align}	
To derive this result we replaced the sum over the random errors by their RMS values, giving a better sense of the relative strengths of the perturbations and the desired interactions (see Appendix~\ref{app:B} for details of the calculation).  As Eq.~\eqref{eq:hnewB} demonstrates, having multiple copies of the model $H_Q$ boosts its energy scale by a factor of $K$ (the $KH_Q$ term), while due to the assumption of independent errors, the energy scale of the error Hamiltonian grows by only a factor of $\sqrt{K}$.
Thus, relative to the code space gap, the \emph{effective noise strength has been reduced} by a factor of $\sqrt{K}$.  

\subsection{Example of a protected QUBO instance}
\label{sub:example_qubo_instance}
\begin{figure}[t]
	\begin{center}
	\includegraphics[width=\columnwidth]{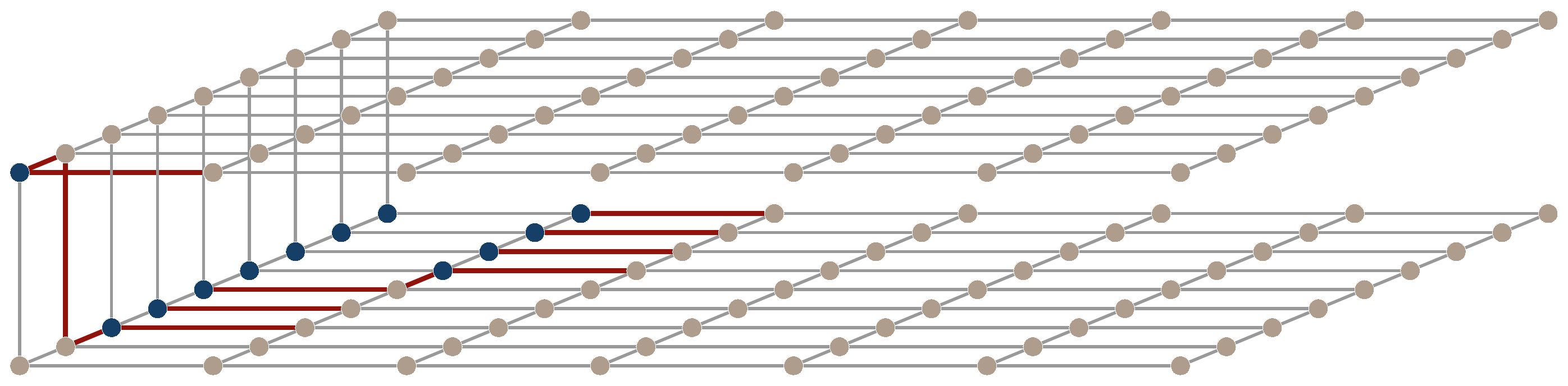}\vspace{.3cm}
	\includegraphics[width=\columnwidth]{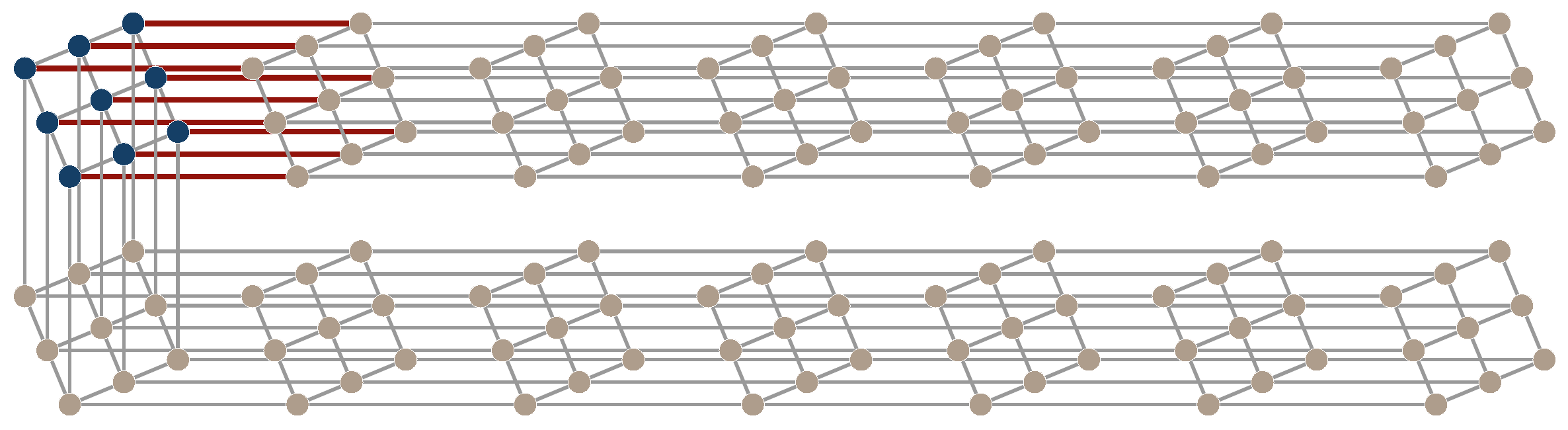}\vspace{.3cm}
	\includegraphics[width=\columnwidth]{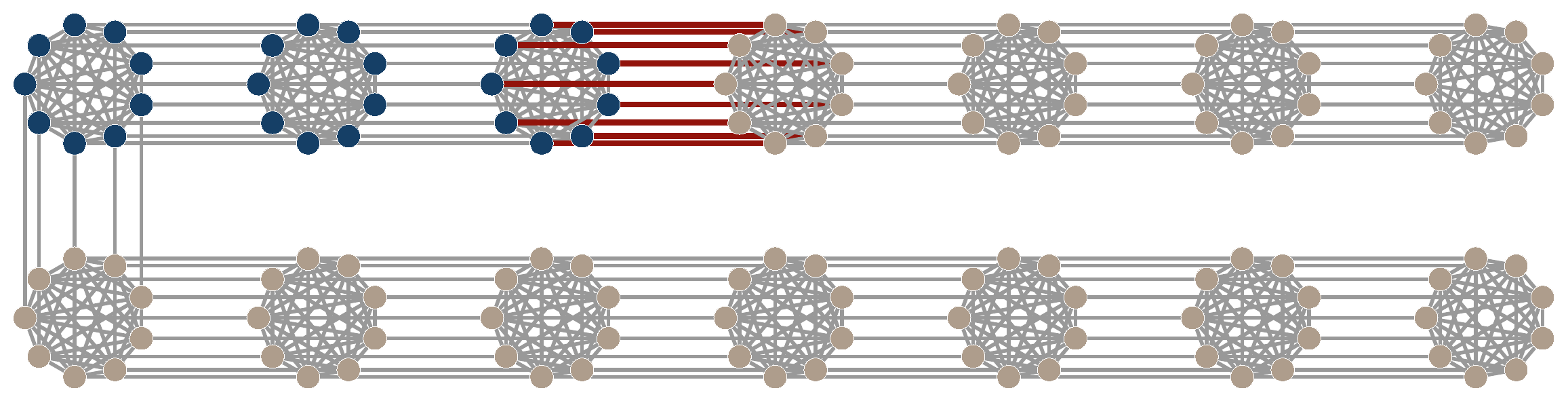}
	\caption{QUBO problem with various repetition encodings.  Violated links are indicated by bold, red lines.\\
	(Top) \textbf{9-bit linear encoding} 
		Errors are slightly reduced in frequency as compared to unencoded problem, 
		but faults are likely to lie outside of the code space, as shown.\\
	(Middle) \textbf{9-bit grid encoding} 
		Errors are significantly less likely to occur than in the unencoded problem.  
		All observed errors were in the code space, \emph{i.e.}, all physical qubits within each 
			logical qubit are in the same state. \\
	(Bottom) \textbf{9-bit complete encoding} 
		Within each logical qubit, all physical qubits are ferromagnetically linked to each other.
		Errors are significantly less likely to occur than in the unencoded problem, but results were not 
			significantly better than the simple 2D grid.  
		All observed errors were in the code space.
		For clarity, erred couplings are not shown. }
	\label{fig:protected}
	\end{center}
\end{figure}

\noindent 
To demonstrate the benefit of the encoding we consider the perturbed model shown in Fig.~\ref{fig:error} (bottom) and protect it by encoding in three different repetition codes: (i) a linear chain of 9 qubits, (ii) a 3$\times$3 grid, (iii) the complete graph of 9 bits.  The linear chain does not satisfy the heuristic that the degree of the encoding graph be larger than that of the problem graph, so errors are not likely to be in the code space.  Our simulation data confirm this heuristic; the ground state usually lies outside the code space.  Cases (ii) and (iii), on the other hand, each satisfy this heuristic. Their performance is approximately equivalent, and failing instances are all seen to lie in the code space.  All three encodings are shown in Fig.~\ref{fig:protected}.

\begin{figure}[h!]
	\begin{center}
	\includegraphics[width=\columnwidth]{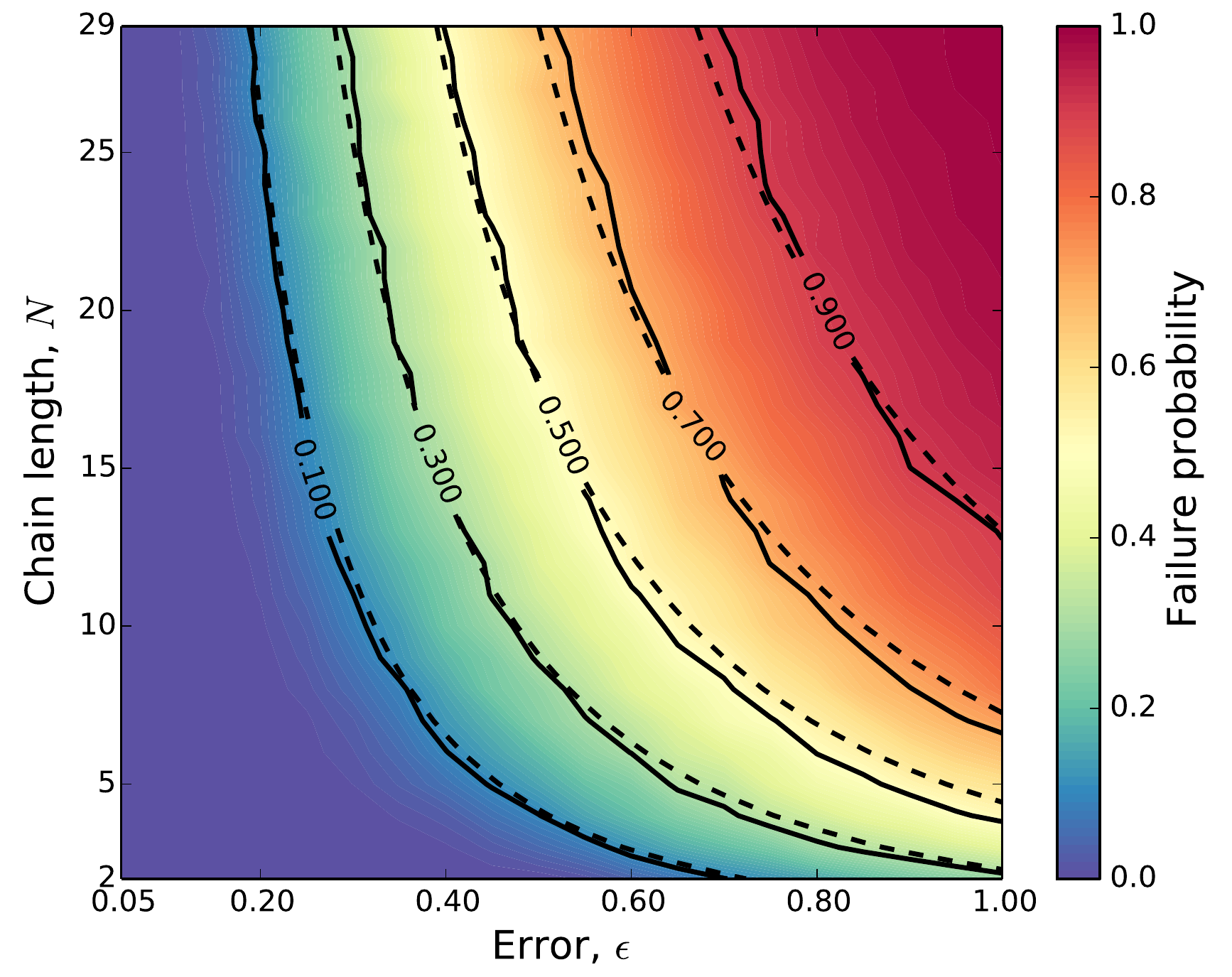}\vspace{.3cm}
	\includegraphics[width=\columnwidth]{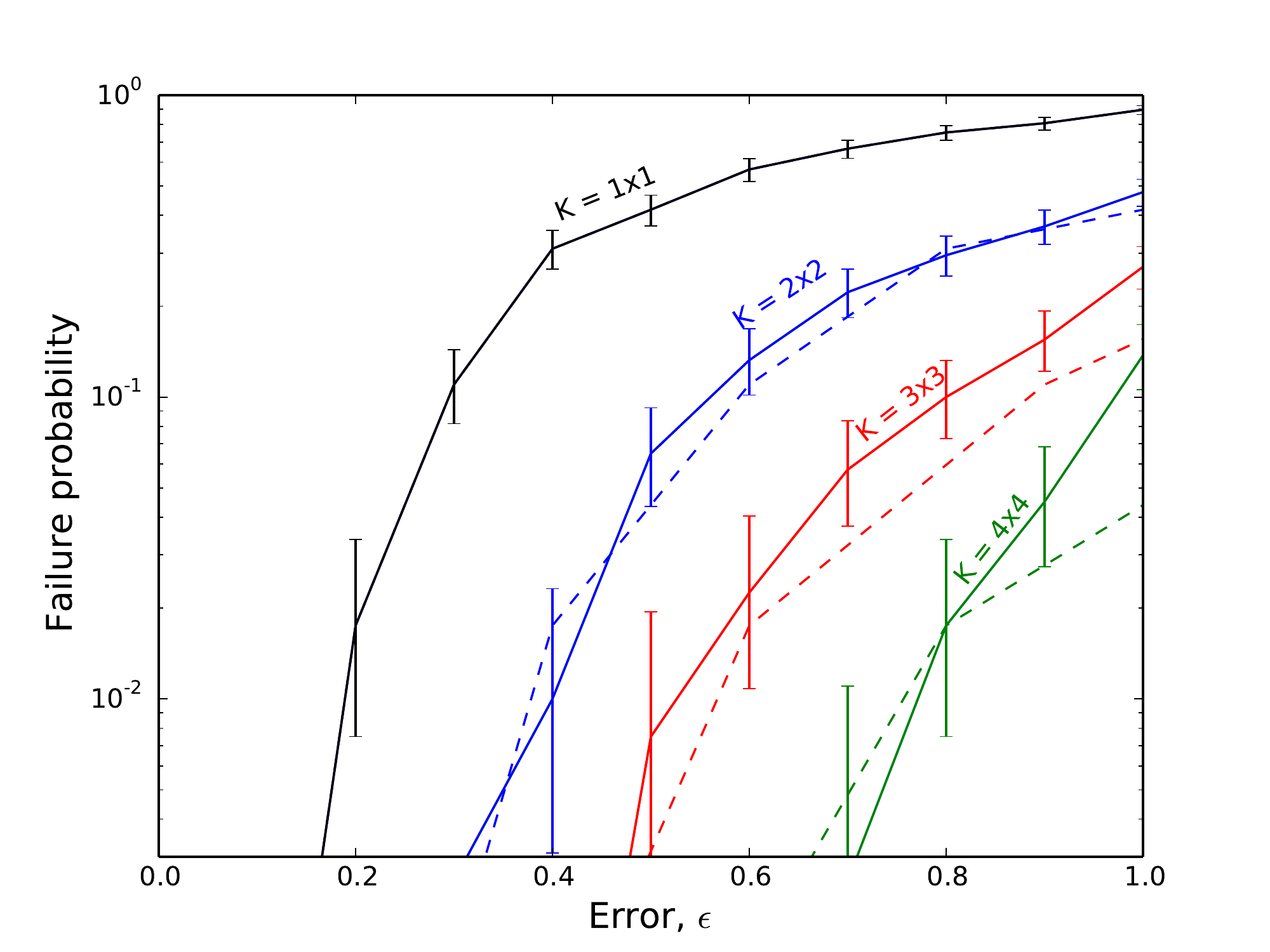}
	\caption{Failure probability of Ising chain.\\  
	(Top) \textbf{Unencoded failure probabilities}
		Color contour plot of the unencoded failure probabilities for a variable 
		length Ising ladder at $\epsilon\in[0,1]$.  Also shown (dashed, unlabeled) are contours of $\sqrt{N}\epsilon$.  
		The close overlap of these with the (solid, labeled) contours of the failure probability
	 	implies that the failure probability is a function of $\sqrt{N}\epsilon$, as expected. 
		Computed with 1000 MC iterations per datapoint.\\
	(Bottom) \textbf{Encoded failure probabilities} 
		(solid) Failure probabilities for an Ising ladder of length $N=13$ encoded in four different 2D grid repetition codes.  (dashed) Failure probability of unencoded system with epsilon reduced by a factor of $\sqrt{K}$.  The tendency to disagree at high epsilon can be attributed to an increased probability of failures lying outside the code space as $\epsilon$ approaches $E_{\rm max}=1$. 
		These data were computed with 400 MC iterations per datapoint.  The error bars indicate uncertainty in the failure probabilities resulting from the finite MC sample set.  }
	\label{fig:protected_detailed}
	\end{center}
\end{figure}

In Fig.~\ref{fig:protected_detailed} we further explore the parameter space by computing (1000 MC iterations per datapoint) the failure probabilities as a function of both the chain length, $N$, and the strength of the errors.  Also shown in the top subfigure are the (irregularly spaced) contours of $\sqrt{N}\epsilon_{\rms}$.  The close overlap of these contours with the failure probability contours is as expected from the energetic arguments above.  The bottom figure demonstrates the $\sqrt{K}$ reduction in the failure probability due to encoding.  The deviations at large $\epsilon_{\rm max}$ are due to failures in the error model assumptions. 

\section{Formulation in terms of stabilizer codes} 
\label{sec:formulation_in_terms_of_stabilizer_codes}
\noindent
The repetition encoding described above may be considered as a special case of more general stabilizer encodings \cite{Gottesman:1996fk,Gaitan:book,Lidar-Brun:book}.  Recall that quantum stabilizer codes function by distributing the information stored in a small quantum system over a much larger quantum system.  In this way, local perturbations are no longer sufficient to destroy the encoded state.  Measurements of the \emph{stabilizer generators} are sufficient to detect (and possibly correct) errors, without causing measurement-induced decoherence of the encoded state.  This ability to detect errors has been the primary use of quantum stabilizer codes in both digital and analog QIP, most notably by enabling the circuit model fault-tolerance theorem.  However, generalizing \cite{PAL:13}, here we leverage an additional facet of stabilizer codes: that operators on the encoded information may be implemented in many ways.  That this technique increases the success probability of the algorithm follows from two observations:
\begin{enumerate}
	\item Within the code space, $K$-fold multiplicity of the logical operators increases the energy scale by a factor of $K$, while increasing the noise strength by a factor of only approximately $\sqrt{K}$.
	\item Adding the stabilizer generators, with negative weights, to the Hamiltonian energetically favors the code space by penalizing errors detected by the code \cite{jordan2006error}.
\end{enumerate}

For the repetition code discussed above, the stabilizer group is generated by two-qubit parity checks (Pauli $ZZ$ operators) within each logical qubit, 
\begin{equation*}
\mathcal{S} = \left\langle \{Z_{i,1}Z_{i,k} \vert k\in[2\to K], i\in[1\to N]\}\right\rangle.
\end{equation*} 
A valid set of logical operators is $\overline X_i = \otimes_{k=1}^K X_{i,k}$ and $\overline Z_i = Z_{i,k}$, with $k\in\{1,\dots,K\}$.  The two-qubit logical operators may then be formed as $\overline{Z_iZ_j} = Z_{i,k}Z_{j,k}$.
For each two-qubit logical operator appearing in the unencoded Hamiltonian, our encoding uses $K$ of these equivalent logical operators, which increases the code space energy scale (and thus the protection against fluctuations, which are uncorrelated, add incoherently, and thus scale only as $\sqrt{K}$) by a factor of $K$.  The addition of the stabilizers, which here take the form of ferromagnetic links within each logical qubit, force the ground state to lie in the code space.  Stacking even more equivalent logical operators would work even better, but although many more equivalent logical operators exist, most of them will be high-weight and/or will involve physically distant qubits that cannot feasibly be coupled together.

This redundancy in the definition of the logical operators extends to \emph{any} stabilizer code for \emph{any} analog QIP. To see this, consider a generic Hamiltonian on $N$ qubits,
\begin{equation}
\label{eq:unencHam}
	H = \sum_i \alpha_i L_i,
\end{equation}
where $L_i$ is some operator.  Encoding the system with a quantum stabilizer code expands the Hilbert space to $M>N$ physical qubits (e.g., $M=NK$ in the discussion above, where $N$ is now the number of encoded qubits).  Let $\overline{L_i}$ denote any valid encoding of $L_i$ using the given stabilizer code. Because this code possesses $M-N$ stabilizer generators, $\{S_j\}_{j=1}^{M-N}$, each logical operator may be encoded in any of $2^{M-N}$ different ways,
\begin{equation}
\label{eq:logicals}
	\overline{L_{i,\mathbf{s}}} = \overline{L_i} \prod_{j=1}^{M-N} S_j^{s_j}
\end{equation}
where $\mathbf{s}=\{s_1,\dots,s_{M-N}\}$ is a binary vector indexing an element of the stabilizer group, and $S_j^0\equiv \ident$.  We can then replace each logical operator in the Hamiltonian by a sum over these equivalent logical operators,
\begin{equation}
\label{eq:newHam}
	H \mapsto \sum_i \alpha_i \sum_{\mathbf s} \beta_{i\mathbf{s}} \overline{L_{i,\mathbf{s}}},
\end{equation}
with the constraint that the quantity $\sum_{\mathbf{s}} \beta_{i\mathbf{s}} = K$ is a constant which is  independent of $i$. This scales the logical Hamiltonian strength by a factor of $K$. Of course, the best we can hope to achieve is to implement \emph{all} $2^{M-N}$  equivalent logical operators. However, many of these operators will involve qubits which are spatially very far from one another.  Engineering constraints will ultimately limit which of the equivalent logical operators are physically realizable.  We assume that any logical operator which can be implemented is indeed implemented.  As before, while this encoding increases the number of error terms which can affect the system, the expected impact of these errors grows only as $\sqrt{K}$.  Finally, adding terms in the stabilizer group to the Hamiltonian penalizes states outside the ground space.

Unfortunately, many quantum codes possess high-weight logical or stabilizer operators, so are very difficult to implement physically. For example, the logical $X$ operator of the repetition code has weight $K$, which precludes an encoding of the initial transverse field Hamiltonian $H(0) = -\sum_i X_i$.  Nonetheless, for the classical QUBO problem we considered, the repetition code is particularly suitable for an encoding of the problem Hamiltonian because it requires only weight-\emph{two} operators, which is also the reason that it was the code of choice in the experimental work of \cite{PAL:13}. It is an interesting open problem to come up with an encoding which consists only of weight-two logical operators, or to prove that no such encoding exists \cite{jordan2006error}.

\section{Discussion} 
\label{sec:discussion}
\noindent
The encoding described above forces the systems to adopt the ground state configuration of a noisy Hamiltonian \emph{averaged over $K$ instances of the noise}.  This averaging reduces the effective noise strength by a factor of $\sqrt{K}$, greatly increasing the probability of finding the ground state of the correct problem. The penalty term is crucial, since had the copies been uncoupled from each other, each would be free to choose its own ground state, a problem that would be exacerbated even in the absence of control noise if the ground state is multiply-degenerate. 

The key assumption behind this approach and the results presented here is that the errors are independent and identically distributed.  This assumption is used when computing the RMS expectation value of the sum of the perturbations.  If the perturbations are correlated, then not all the cross terms vanish.  Even so, some suppression is expected for correlated errors, except in the event that the errors are perfectly correlated, for then the expectation value of the sum is simply $K \epsilon_\rms$, which shows no improvement over the unencoded case.  However, this case can be viewed as a systematic error, which can be eliminated using better engineering. Thus we expect that the approach presented here will prove to be beneficial under quite general conditions. Indeed, the experimental results presented in \cite{PAL:13}, where a $3$-qubit repetition code was used mainly to protect against environmental noise, demonstrate the versatility of the encoding technique presented here. The key challenge is to extend these ideas into a full-fledged fault tolerance theorem for analog quantum information processing.


\acknowledgments
\label{sec:acknowledgements}
\noindent
The authors thank Alexey Gorshkov for calling attention to the problem of control errors on the final Hamiltonian in adiabatic quantum computing at the 2013 KITP Workshop on Control of Complex Quantum Systems. Sandia National Laboratories is a multi-program laboratory managed and operated by Sandia Corporation, a wholly owned subsidiary of Lockheed Martin Corporation, for the U.S. Department of Energy's National Nuclear Security Administration under contract DE-AC04-94AL85000. DAL's research was supported by ARO-MURI grant W911NF-11-1-0268, by ARO-QA grant number W911NF-12-1-0523, by the Lockheed Martin Corporation, and by NSF grant numbers PHY-969969 and PHY-803304.
\bibliography{wrong_ham-2}

\begin{appendix}
\section{Encoding as Weighted MAX-2-SAT} 
\label{sec:encoding_as_max_sat}
\noindent
Finding the minimum eigenvalue of the (fully classical) QUBO Hamiltonian may be easily mapped to a weighted MAX-2-SAT instance and solved with available solvers.  The QUBO problem may be specified as 
\begin{align}
	\underset{\vec Z}{\rm minimize} 
		\notag
		&\;\; \sum_{ij} \frac{1}{2} A_{ij} Z_i Z_j + \sum_i a_i Z_i \\
		\label{eq:QUBOmin}
	{\rm subject\;to} 
		&\;\; Z_i \in \{\pm 1\}
\end{align}
where the coupling matrix $A$ has been assumed symmetric with diagonal entries equal to zero (\emph{i.e.,} is hollow).  The MAX-2-SAT problem, on the other hand, is to determine the bit string \emph{maximizing} the sum of the weights of satisfied boolean clauses,
\begin{align}
	\underset{\vec X}{\rm maximize} 
		\notag
		&\;\; \sum_{ij} \frac{1}{2} B_{ij} X_i \vee X_j + \sum_i b_i X_i \\
	{\rm subject\;to} 
		\label{eq:MAX2SAT}
		&\;\; X_i \in \{0,1\}
\end{align}
Here again $B$ is symmetric and hollow.  We may convert between them by assigning  $Z_i \mapsto (1-2 X_i)$ and noting that
\begin{equation}
\label{eq:ortotimes}
	X_i \vee X_j = X_i + X_j - X_i X_j.
\end{equation}
After a bit of algebra, we see that our QUBO problem may be stated as the following MAX-2-SAT problem:
\begin{align}
	\underset{\vec X}{\rm maximize} 
		\notag
		&\;\; \sum_{ij} \frac{1}{2}A_{ij} X_i \vee X_j + \sum_i \left(a_i - \sum_j A_{ij} \right) X_i \\
	{\rm subject\;to} 
		\label{eq:MAX2SAT}
		&\;\; X_i \in \{0,1\} .
\end{align}

\section{Derivation of Eq.~\eqref{eq:hnewB}}
\label{app:B}
To derive Eq.~\eqref{eq:hnewB} we proceed as follows, making use of Eq.~\eqref{eq:PZP}, and where to arrive at Eq.~\eqref{eq:RMS} we replace the sum over the random errors by their RMS values:
\begin{widetext}
\bes
\begin{align}
\label{eq:hnewBfull}
P\overline{H^\prime}P 
			=&\; P \left[ \sum_{\left\langle i,j \right\rangle \in \mathcal{E},k\in \mathcal{V}_F} 
				\left(J_{ij}+\epsilon_{ij,k}^{(J)}\right) Z_i Z_j
				+ \sum_{i\in\mathcal{V},k\in\mathcal{V}_F} \left(h_i+\epsilon^{(h)}_{i,k}\right) Z_i 
				+ \sum_{i\in\mathcal{V}}\sum_{\left\langle k,l \right\rangle \in \mathcal{E}_{F}} (-J_F +\epsilon_{i,kl}^{(J)})Z_i Z_i \right] P \\
			=&\; P \left[  \sum_{\left\langle i,j \right\rangle \in \mathcal{E}}  
				\left( K J_{ij}+\sum_{k\in\mathcal{V}_F} \epsilon_{ij,k}^{(J)} \right)Z_i Z_j 
				+ \sum_{i\in\mathcal{V}} \left(K h_i+\sum_{k\in\mathcal{V}_F} \epsilon^{(h)}_{i,k}\right) Z_i 
				-J_F \abs{\mathcal{V}} \abs{\mathcal{E}_{F}}\ident  
				+ \sum_{i\in\mathcal{V}}\sum_{\left\langle k,l \right\rangle \in \mathcal{E}_{F}}\epsilon_{i,kl}^{(J)}\ident
				\right] P \\
				\label{eq:RMS}
	  \simeq &\; P \left[  \sum_{\left\langle i,j \right\rangle \in \mathcal{E}}  
				\left( K J_{ij}+ \sqrt{K} \epsilon^{(J)}_\rms \right)Z_i Z_j 
			 	+ \sum_{i\in\mathcal{V}} \left(K h_i+\sqrt{K} \epsilon^{(h)}_\rms\right) Z_i 
				-\left(J_F \abs{\mathcal{V}}\abs{\mathcal{E}_F} - \sqrt{ \abs{\mathcal{V}}\abs{\mathcal{E}_F}}\epsilon^{(J)}_\rms\right)\ident
				\right] P  \\
				\label{eq:suppress}
			=& P \left[  K H_Q + \sqrt{K}\left(\sum_{\left\langle i,j \right\rangle \in \mathcal{E}}  
				 \epsilon^{(J)}_\rms Z_i Z_j 
			 	+ \sum_{i\in\mathcal{V}} \epsilon^{(h)}_\rms Z_i \right) +\left(-J_F \abs{\mathcal{V}}\abs{\mathcal{E}_F} + \sqrt{ \abs{\mathcal{V}}\abs{\mathcal{E}_F}}\epsilon^{(J)}_\rms\right)\ident
				\right] P .
	\end{align}	
\ees
\end{widetext}

\end{appendix}

\end{document}